\documentclass{webofc}

\usepackage[varg]{txfonts}   
\usepackage{hyperref}
\usepackage{url}

\hypersetup{colorlinks=true,citecolor=blue,urlcolor=blue,linkcolor=blue}
\begin{document}
\title{Listening to the long ringdown}
%
%
\subtitle{A novel way to pinpoint the EOS in neutron-star cores}

\author{\firstname{Christian} \lastname{Ecker}\inst{1}\fnsep\thanks{\email{ecker@itp.uni-frankfurt.de}}
\and
\firstname{Tyler} \lastname{Gorda}\inst{2}
\and
\firstname{Aleksi} \lastname{Kurkela}\inst{3}
\and
\firstname{Luciano} \lastname{Rezzolla}\inst{1}
}

\institute{
Institut f\"ur Theoretische Physik, Goethe Universit\"at, Max-von-Laue-Str. 1, 60438, Germany
\and
Department of Physics, The Ohio State University, Columbus, OH 43210, USA
\and
Faculty of Science and Technology, University of Stavanger, Stavanger, 4036, Stavanger, Norway
}

\abstract{
Gravitational waves (GWs) from binary neutron star (BNS) merger remnants complement constraints from the inspiral phase, mass–radius measurements, and microscopic theory by providing information about the neutron-star equation of state (EOS) at extreme densities.
We perform general-relativistic simulations of BNS mergers using EOS models that span the uncertain high-density regime. 
We find a robust correlation between the ratio of energy and angular momentum lost during the late-time post-merger GW signal—the long ringdown—and the EOS at the highest densities in neutron star cores.
Applying this correlation to post-merger GW signals reduces EOS uncertainty at several times saturation density, where no direct constraints currently exist.
}
\maketitle
\section{Introduction}
\label{intro}
The densest matter in the observable universe resides in neutron star (NS) cores, where gravity compresses matter to densities far exceeding nuclear saturation, $n_{\rm sat} = 0.16$ baryons/fm$^3$. The equation of state (EOS) governing such strongly interacting matter remains uncertain, yet its precise determination could provide important insights into the phase diagram of Quantum Chromodynamics (QCD).
Recent progress in constraining the EOS has come from both improved NS observations and ab-initio theory (e.g.,~\cite{Annala:2023cwx}). 
Notably, the gravitational-wave (GW) signal from the BNS merger GW170817 revealed that tidal deformabilities of inspiralling NSs—closely linked to the EOS around $3~n_\mathrm{sat}$—can be measured (e.g.,~\cite{Radice:2020ddv}).
Third-generation GW observatories~\cite{Evans:2021gyd,Abac:2025saz}. will have the sensitivity to detect post-merger signals with high SNR, providing access to densities even higher than those in the pre-merger phase.
These signals exhibit spectral features that are tightly correlated with the EOS (see~\cite{Huez:2025gja} and references therein).
In recent work~\cite{Ecker:2024uqv}, we propose a refined approach that targets a specific late-time component of the post-merger signal—occurring between $1$ and $15$ ms—which we term the long ringdown. We briefly summarize this in the present proceedings contribution (for details see~\cite{Ecker:2024uqv}).
Analogous to black hole ringdowns, this slowly damped signal arises from a quasi-stationary, nearly axisymmetric hypermassive neutron star emitting nearly monochromatic GWs. We show that during this stage, the emitted GW energy and angular momentum follow a linear relation tied directly to the high-density EOS. 
Thus, observing the long ringdown at high SNR enables direct EOS constraints at the highest observable densities.

\section{Methods}
\label{sec-1}
For the agnostic EOS construction, we use the Gaussian Process framework presented in~\cite{Gorda:2022jvk}. To efficiently represent the EOS with a limited number of samples, we consider the posterior in the 4D parameter space $(M_\text{TOV},\,C_\text{TOV}, \, \ln p_\text{TOV}, \, R_{1.4})$, within which we perform a principal component analysis to identify a small set of EOSs that characterize the $68\%$ credible region of the distribution. From this analysis, we select six “golden” EOSs from our ensemble: EOSs with the highest likelihood near the extrema of the $68\%$ credible region, plus one near the origin.

We use the FUKA~\cite{Papenfort:2021hod} and FIL~\cite{Most:2019kfe} codes to simulate BNS mergers with parameters consistent with GW170817, i.e., with
fixed chirp mass $\mathcal{M}_{\rm chirp} = 1.18\,M_\odot$ and three different ratios $q := M_2 / M_1 =0.7,0.85,1$ of the binary constituent masses $M_1$ and $M_2$.
From these simulations, we extract the GW strain components ($h_+$,$h_\times$) and compute the slope of the emitted energy ($E_{\rm GW}$) with respect to the angular momentum ($J_{\rm GW}$), as well as the instantaneous frequency ($f_{\rm GW}$)
\begin{equation}
  \label{eq:fGWvsdEJ}
  \frac{dE_{\rm GW}}{dJ_{\rm GW}}=\frac{\dot{E}_{\rm GW}}{\dot{J}_{\rm GW}}=\frac{\dot{h}_+^2+\dot{h}_\times^2}{h_+\dot{h}_\times-\dot{h}_+h_\times}\,,
  \qquad
  f_{\rm GW}=\frac{1}{2\pi}\frac{h_+\dot{h}_\times-\dot{h}_+h_\times}{h_+^2+h_\times^2}\,.
\end{equation}
For a simple system with an $\ell=2, m=2$ deformation, one finds the identity ${\dot{E}_{\rm GW}}/{\dot{J}_{\rm GW}}=f_{\rm GW} / (2\pi)$, implying a linear slope during the long ringdown, where $f_{\rm GW}(t) \simeq {\rm const.}$.
In practice we normalize them by their values at merger-time $\hat{E}_{\rm GW}=E_{\rm GW}/E^{\rm mer}_{\rm GW}$,$\hat{J}_{\rm GW}=J_{\rm GW}/J^{\rm mer}_{\rm GW}$.

\section{Results}
\label{sec-2}
\begin{figure}[h]
\centering
\includegraphics[height=0.33\textwidth,clip]{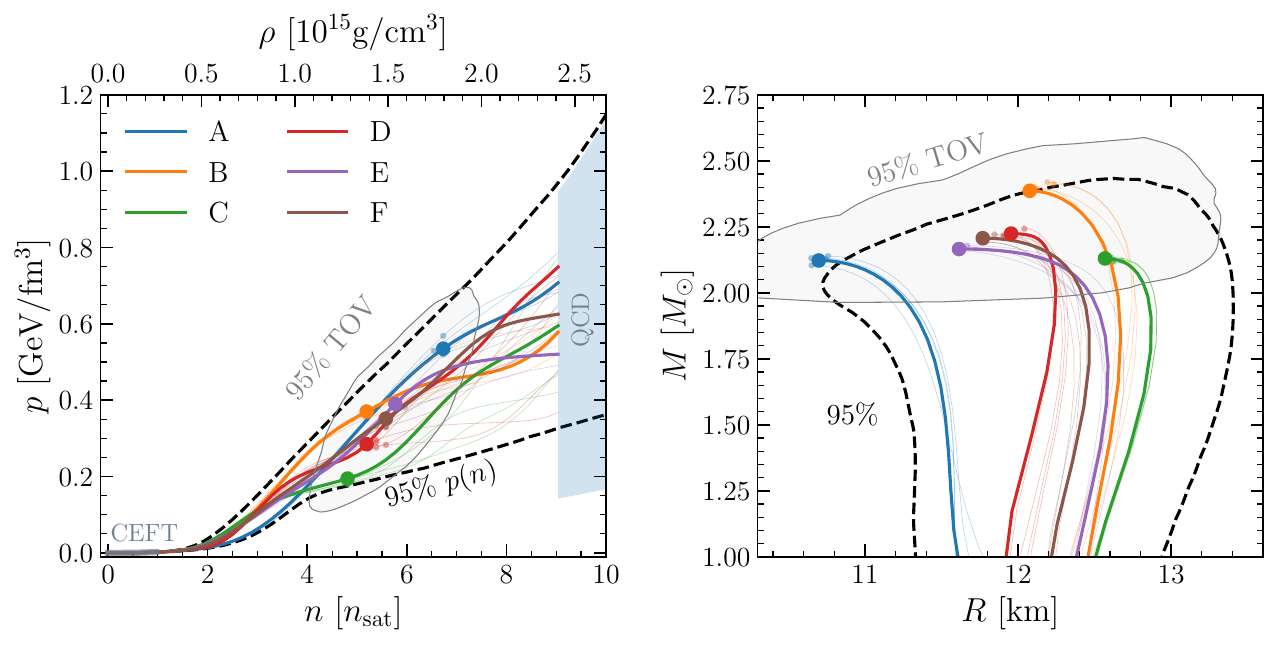}\,\,\includegraphics[height=0.33\textwidth,clip]{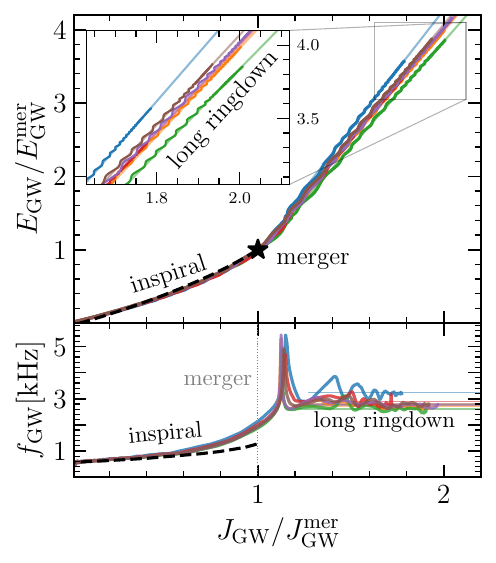}
\caption{
 \textit{Left panel:} Solid lines of different colors show the six golden EOSs (${\rm A}$--${\rm F}$) in the $(p,n)$ plane. 
 The dashed black lines show the $95\%$ credible intervals of all possible EOSs, while the CEFT and QCD bounds are shown with shaded areas. 
 Colored filled circles show the TOV points of the golden EOSs, while the solid light gray line is the $95\%$ credible interval for all TOVs.  
 \textit{Middle panel:} The same as in the left but shown in the $(M,R)$ plane.
 \textit{Right panel:} Radiated GW energy and angular momentum (top), normalized to their merger values, and instantaneous GW frequency $f_\mathrm{GW}$ (bottom).
}
\label{fig-1}
\end{figure}

In Figure~\ref{fig-1} (left panel), the thick colored lines show our selection of golden EOS models, with their corresponding ten closest neighbours in the 4D space plotted as thin lines; dots indicate the central densities reached in maximally massive stars. 
The black dashed lines denote the $95\%$ confidence interval of the entire EOS ensemble, while the blue regions indicate the imposed uncertainty bands from chiral effective field theory (CEFT) and perturbative QCD. The middle panel presents the resulting mass–radius curves obtained by solving the Tolman–Oppenheimer–Volkoff (TOV) equations. The right panel displays one of our main results: the relation between the normalized GW energy and angular momentum (top), together with the corresponding instantaneous GW frequency (bottom), as obtained from numerical BNS merger simulations—here shown for equal-mass binaries.
The key insight from these results is that $d\hat{E}_{\rm GW}/d\hat{J}_{\rm GW}=$const. correlates with the properties of the golden EOSs. 

\begin{figure}[h]
\centering
\includegraphics[width=0.6\textwidth,clip]{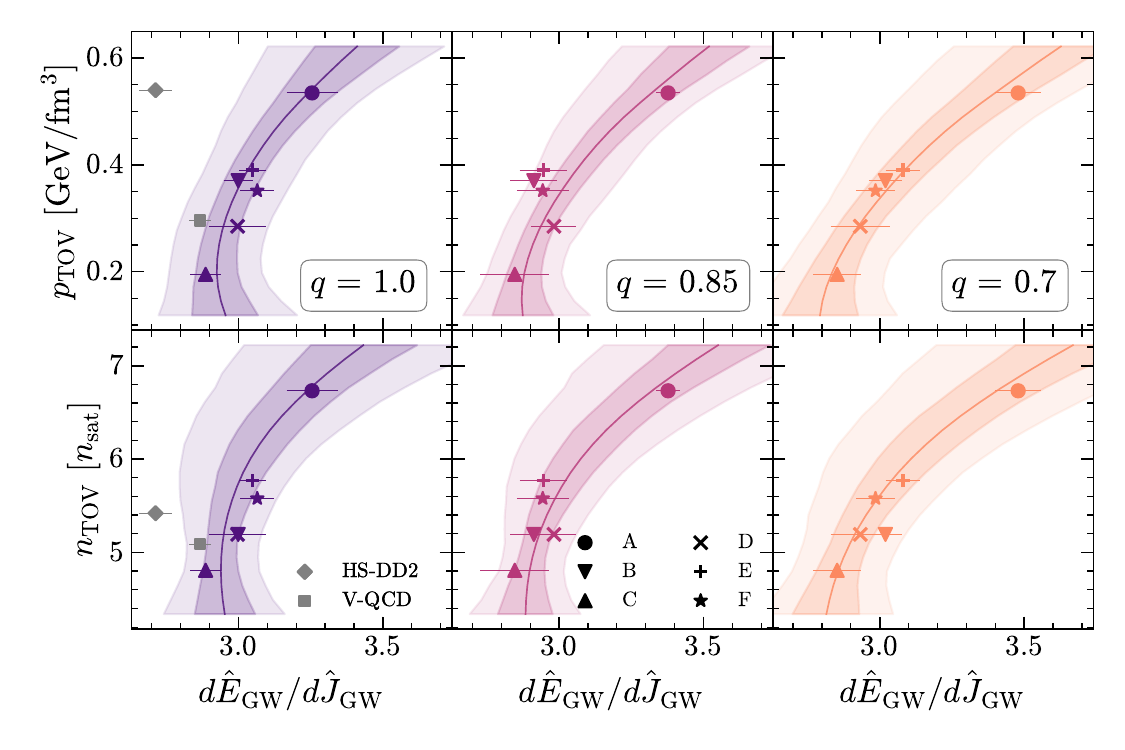}
\caption{
 Data and fits of the correlations between the slope $d\hat{E}_{\rm GW}/d\hat{J}_{\rm GW}$ and $p_\mathrm{TOV}$ (top row) or $n_\mathrm{TOV}$ (bottom row), and for different mass ratios (different columns).
 The dark (light) shaded regions denote $68\%(95\%)$ credible intervals for the bilinear fitting model, while the solid lines denote the mean value.
}
\label{fig-2}
\end{figure}

In Fig.~\ref{fig-2}, the shaded regions show bilinear fits to the correlations between $d\hat{E}_{\rm GW}/d\hat{J}_{\rm GW}$, extracted from our simulations (symbols), and the central pressure (top) and number density (bottom) of maximally massive neutron stars (details in figure caption). 

\begin{figure}[h]
\centering
\includegraphics[width=0.65\textwidth,clip]{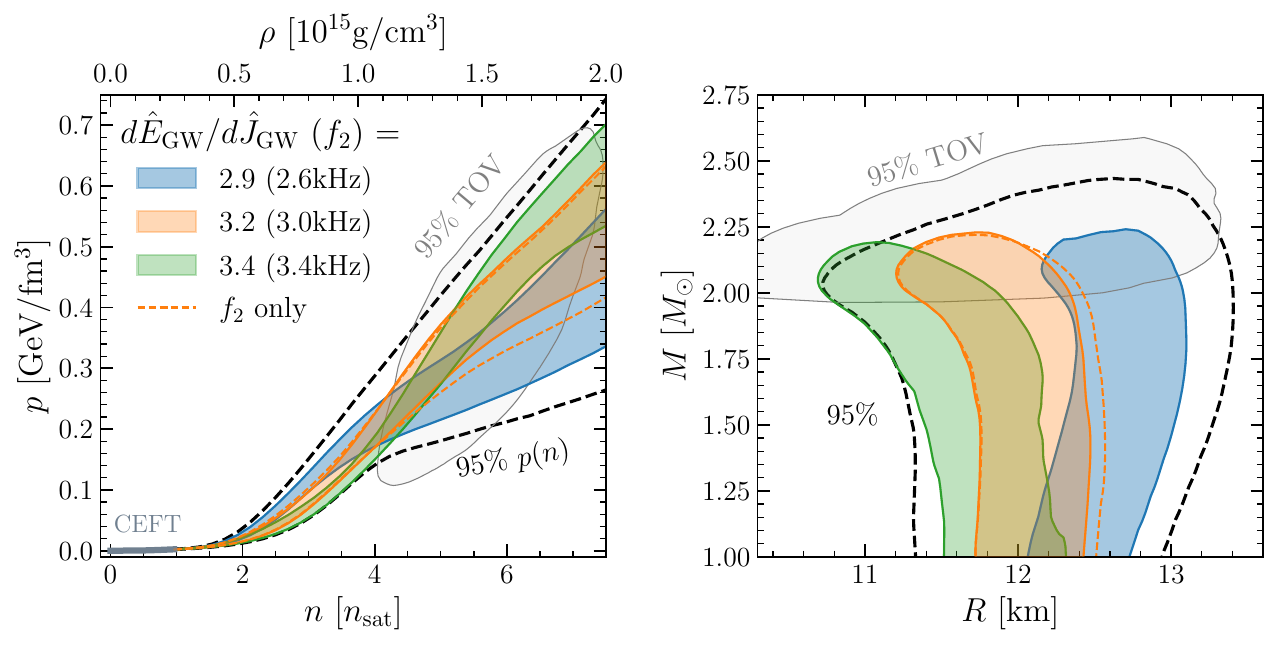}
\caption{ 
 \textit{Left panel} shows the $68\%$ credibility values on the $(p,n)$ plane from potential \textit{joint} measurements of the slope and $f_2$ with $\pm 4\%$ uncertainty.
  \textit{Right panel} the same but in the $(M,R)$ plane.
  }
\label{fig-3}
\end{figure}

These correlations in conjunction with a future post-merger GW detection, can be used to constrain the EOS at \textit{all densities}.
More specifically, we use as an additional component of the likelihood function $P(\mathrm{data} | \mathrm{EOS}, \mathrm{NSs})$ the integral of our bilinear model over the measurement likelihood, in order to infer the combined EOS and NS properties via Bayes’s theorem
\begin{equation}
    P(\mathrm{EOS}, \mathrm{NSs} | \mathrm{data}) = \frac{P(\mathrm{data} | \mathrm{EOS}, \mathrm{NSs}) P(\mathrm{EOS}, \mathrm{NSs})}{P(\mathrm{data})}\,.
\end{equation}
This is illustrated in Fig.~\ref{fig-3}, which presents the results of mock measurements for three different slope values and $f_2$, assuming a $\pm4\%$ measurement uncertainty.
Our results show that such potential future measurements could lead to a significant reduction of the EOS uncertainty (left panel) and the corresponding mass-radius distribution of isolated neutron stars (right panel).

\section{Summary}
\label{sec-3}
Third-generation gravitational-wave observatories will detect numerous BNS post-merger signals with high signal-to-noise ratios, enabling detailed studies of dense matter. We explore correlations between gravitational-wave observables and the equation of state (EOS) using a large set of physically consistent, generic EOSs. A principal component analysis identifies a reduced set of representative “golden” EOSs that capture the main variation in post-merger dynamics.
We discover a novel correlation between the slope of the long ringdown phase—quantified by $d\hat{E}_{\rm GW}/d\hat{J}_{\rm GW}$—and the pressure and density at the maximum-mass Tolman–Oppenheimer–Volkoff (TOV) configuration, $(p_{\rm TOV}, n_{\rm TOV})$. This slope is straightforward to extract from the waveform and provides direct constraints on the EOS at the highest core densities.
By combining this correlation with Bayes’ theorem, we obtain improved constraints on the mass–radius relation across the full density range, outperforming methods based solely on the post-merger peak frequency $f_2$.

\end{document}